\documentclass{article}
\usepackage{natbib,epsfig,emulateapj}
\begin{document}

\lefthead{Andersson et al.}
\righthead{R-mode runaway and rapidly rotating neutron stars}

\title{R-mode runaway and rapidly rotating neutron stars} 

\author{Nils Andersson and David Ian Jones} \affil{Department of
  Mathematics, University of Southampton, Southampton SO17 1BJ, United
  Kingdom} 
\centerline{\it \small na@maths.soton.ac.uk, dij@maths.soton.ac.uk}
\vspace{0.2cm}

%\email{na@maths.soton.ac.uk, dij@maths.soton.ac.uk}

\author{Kostas D. Kokkotas and Nikolaos Stergioulas} \affil{Department
  of Physics, Aristotle University of Thessaloniki, Thessaloniki
  54006, Greece} 
\centerline{\it \small kokkotas@astro.auth.gr, niksterg@astro.auth.gr}
\vspace{0.2cm}

\centerline{\it (Accepted March. 10, 2000)}
\vspace{0.2cm}

%\email{kokkotas@astro.auth.gr, niksterg@astro.auth.gr}

\begin{abstract} 
  We present a simple spin evolution model that predicts that rapidly
  rotating accreting neutron stars will mainly be confined to a narrow
  range of spin-frequencies; $P= 1.5-5$~ms. This is in agreement with
  current observations of both neutron stars in the Low-Mass X-ray
  Binaries and millisecond radio pulsars. The main ingredients in the
  model are: i) the instability of r-modes above a critical spin rate,
  ii) thermal runaway due to heat released as viscous damping
  mechanisms counteract the r-mode growth, and iii) a revised estimate
  of the strength of dissipation due to the presence of a viscous
  boundary layer at the base of the crust in an old and relatively
  cold neutron star.  We discuss the gravitational waves that are
  radiated during the brief r-mode driven spin-down phase.  We also
  briefly touch on how the new estimates affects the predicted initial
  spin periods of hot young neutron stars.
\end{abstract}


\section{Introduction} 

The launch of the Rossi X-ray Timing Explorer (RXTE) in 1995 heralded
a new era in our understanding of neutron star physics.  Detailed
observations of quasiperiodic phenomena at kHz frequencies in more
than a dozen Low-Mass X-ray Binaries (LMXB) strongly suggest that
these systems contain rapidly spinning neutron stars (for a recent
review, see \citet{vanderklis00}), providing support for the standard
model for the formation of millisecond pulsars (MSP) via spin-up due
to accretion.

Despite these advances several difficult questions remain to be
answered by further observations and/or theoretical modeling. For
example, we still do not know the reason for the apparent lack of
radio pulsars at shorter periods than the 1.56~ms of PSR1937+21 (for a
review of recent progress in the modelling of rotating neutron stars,
see \citet{stergioulas}).  The recent RXTE observations provide a
further challenge for theorists. Various models suggest that the
neutron stars in LMXB spin rapidly, perhaps in the narrow range
260-590~Hz \citep{vanderklis00}.  Three different models have been
proposed to explain this surprising result.  The first model (due to
\citet{white97}) is based on the standard magnetosphere model for
accretion induced spin-up, while the remaining two models are rather
different, both being based on the idea that gravitational radiation
balances the accretion torque. In the first such model for the LMXB
(proposed by \citet{bildsten98} and recently refined by
\citet{ushomirsky00}), the gravitational waves are due to a quadrupole
deformation induced in the deep neutron star crust because of
accretion generated temperature gradients. The second
gravitational-wave model relies on the recently discovered r-mode
instability (see \citet{akreview} for a review) to dissipate the
accreted angular momentum from the neutron star.

In this Letter we reexamine the idea that gravitational waves from
unstable r-modes provide the agent that balances the accretion torque.
This possibility was first analyzed in detail by \citet{akst99} (but
see also \citet{bildsten98}). Originally, it was thought that an
accreting star in which the r-modes were excited to a significant
level would reach a spin-equilibrium, very much in the vein of
suggestions by \citet{papaloizou} and \citet{wagoner}.  Should this
happen, the neutron stars in LMXB would be prime sources for
detectable gravitational waves.  However, as was pointed out by
\citet{levin99} and \citet{spruit}, the original idea is not viable
since, in addition to generating gravitational waves that dissipate
angular momentum from the system, the r-modes will heat the star up
(via the shear viscosity that counteracts the r-mode at the relevant
temperatures).  Since the shear viscosity gets weaker as the
temperature increases, the mode-heating triggers a thermal runaway and
in a few months the r-mode would spin an accreting neutron star down
to a rather low rotation rate. Essentially, this conclusion rules out
the r-modes in galactic LMXB as a source of detectable gravitational
waves, since they will only radiate for a tiny fraction of the systems
lifetime.

Other recent results would (at first sight) seem to emphasize the
conclusion that the r-modes are not relevant for the LMXB.
\citet{bildsten99} investigated the effect that the presence of a
solid crust would have on the r-mode oscillations.  They estimated
that the dissipation associated with a viscous boundary layer that
arises at the base of the solid crust in a relatively cold neutron
star would greatly exceed that of the standard shear viscosity. Thus,
Bildsten and Ushomirsky concluded that the r-mode instability would
only be relevant for very high rotation rates, and could therefore not
play a role in the LMXB.

We have reassessed the effect of the viscous boundary layer
(correcting an erring factor in the estimates of \citet{bildsten99}).
Our new estimates show that the presence of the crust is important,
but that the instability operates at significantly lower spin rates
than suggested by Bildsten and Ushomirsky.  Once we combine our
estimates with the thermal runaway (now due to heating caused mainly
by the presence of the viscous boundary layer), that results as the
star is spun up to the point at which the instability sets in, we
arrive at a model for the spin-evolution of accreting neutron stars.
Remarkably, this simple model agrees well with existing observations
of rapidly rotating neutron stars, covering both the LMXB and MSP
populations.

\section{Dissipation due to a viscous boundary layer}

The r-mode instability follows after a tug of war between (mainly
current multipole) gravitational radiation that drives the mode and
various dissipation mechanisms that counteract the fluid motion. In
the simplest model, the mode is dominated by shear viscosity at low
temperatures while bulk viscosity may suppress the mode at high
temperatures. At intermediate temperatures, the r-mode sets an upper
limit on the neutron star spin rate.  In an interesting recent paper,
\citet{bildsten99} estimate the strength of dissipation due to the
solid crust of an old neutron star, and find that the presence of a
boundary layer at the base of the crust leads to a very strong damping
of the r-modes.

While we agree with the main idea and the various assumptions made by
Bildsten and Ushomirsky, we would like to point out one important
difference between their results and ones used previously in the
literature. Their assumed timescale for gravitational radiation
reaction differs significantly from, for example, the uniform density
result derived by \citet{kokkotas99} (and subsequently used by several
authors, see \citet{akreview}). This is surprising since the uniform
density result, which can be written
\begin{equation}
t_{gw} \approx  -22 \left( {1.4 M_\odot\over M} \right)
\left( {\mbox{10 km} \over R} \right)^4 \left( {P \over \mbox{1 ms}}
\right)^6 \mbox{ s} \ ,
\label{gwapp}\end{equation}
(where the negative sign indicates that the mode is unstable) has been
shown to be close (within a factor of two) to the results for $n=1$
polytropes. $M$, $R$, and $P$ represent the mass, radius and spin
period of the star, respectively.  In contrast, \citet{bildsten99} use
the $n=1$ polytrope result and argue that it corresponds to $t_{gw}
\approx -146$~s for a canonical neutron star rotating with a period of
1~ms, i.e.  assume that radiation reaction is almost one order of
magnitude weaker than in (\ref{gwapp}).  This difference occurs
because Bildsten and Ushomirsky have only rescaled the fiducial
rotation frequency $\Omega_0=\sqrt{\pi G\bar{\rho}}$ (where
$\bar{\rho}$ represents the average density) in terms of which the
$n=1$ polytrope results of \citet{owen98} were expressed
($t_{gw}\approx -3.26(\Omega_0/\Omega)^6$ for a specific polytropic
stellar model).  Unfortunately, this procedure is not correct.  From
the fundamental relations, e.g. the formula for the gravitational-wave
energy radiated via the current multipoles, one can see that the
gravitational-wave timescale should scale with $M$, $R$ and $P$ in the
way manifested in (\ref{gwapp}). Thus, we believe that Bildsten and
Ushomirsky underestimate the strength of radiation reaction
significantly, which motivates us to reassess the relevance of the
viscous boundary layer.

We should, of course, emphasize at this point that our current
understanding of the r-mode instability is based on crude estimates of
the various timescales.  In order to understand the role of the
instability in an astrophysical context we must improve our modelling
of many aspects of neutron star physics such as the effect of general
relativity on the r-modes, cooling rates, viscosity coefficients,
magnetic fields, potential superfluidity, the formation of a solid
crust etcetera (see \citet{akreview} for a description of recent
progress in these various directions).

In the following we will mainly consider uniform density stars, i.e.
use the gravitational-wave timescale given by (\ref{gwapp}).  In
estimating the dissipation timescale $t_{vbl}$ due to the presence of
the crust, we need to evaluate $t_{vbl} \approx {-2E / (dE/dt)}$ where
$E$ is the mode-energy, and $dE/dt$ follows from an integral over the
surface area at the crust-core boundary (assumed to be located at
radius $R_b$), cf.  equation (3) of \citet{bildsten99}.  To evaluate
this integral we use the standard result for the shear viscosity in a
normal fluid. To incorporate our uniform density model, we make the
reasonable assumption that the density of the star is constant ($\sim
M/R^3$) inside radius $R$. Then it falls off rapidly in such a way
that the base of the crust is located at a radius only slightly larger
than $R$. Hence, it makes sense to use $R_b\approx R$.  If we neglect
the small mass located outside radius $R$ we can then immediately
compare the result for the viscous boundary layer to the timescales
used by \citet{akst99}.  In the end, our estimate for the dissipation
due to the presence of the viscous boundary layer is
\begin{equation}
t_{vbl} \approx 200 
\left( {M \over 1.4 M_\odot } \right)
\left( {\mbox{10 km} \over R } \right)^2 
\left( {T \over 10^8 \mbox{ K}}
\right) \left( {P \over \mbox{1 ms}}
\right)^{1/2} \mbox{ s}\ .
\end{equation}
which is a factor of 2 larger than that of Bildsten and Ushomirsky.
This difference arises simply because the mode-energy $E$ is this
factor larger for uniform density models. The star is assumed to have
a uniform temperature distribution, with core temperature $T$.

In Figure~\ref{fig1} we show the instability window obtained from our
revised estimate. As is clear from this Figure, the presence of a
viscous boundary layer in an old, relatively cold neutron star is,
indeed, important.  However, Bildsten and Ushomirsky's conclusion that
the r-mode instability is irrelevant for the LMXB cannot be drawn from
Figure~\ref{fig1}.  On the contrary, the Figure suggests that the
instability may well be limiting the rotation of these systems.

\section{Thermal runaway in rapidly spinning neutron stars}

\begin{figure*}[t]
\begin{minipage}[t]{3.5in}
\epsfysize=6cm
\centerline{\epsfbox{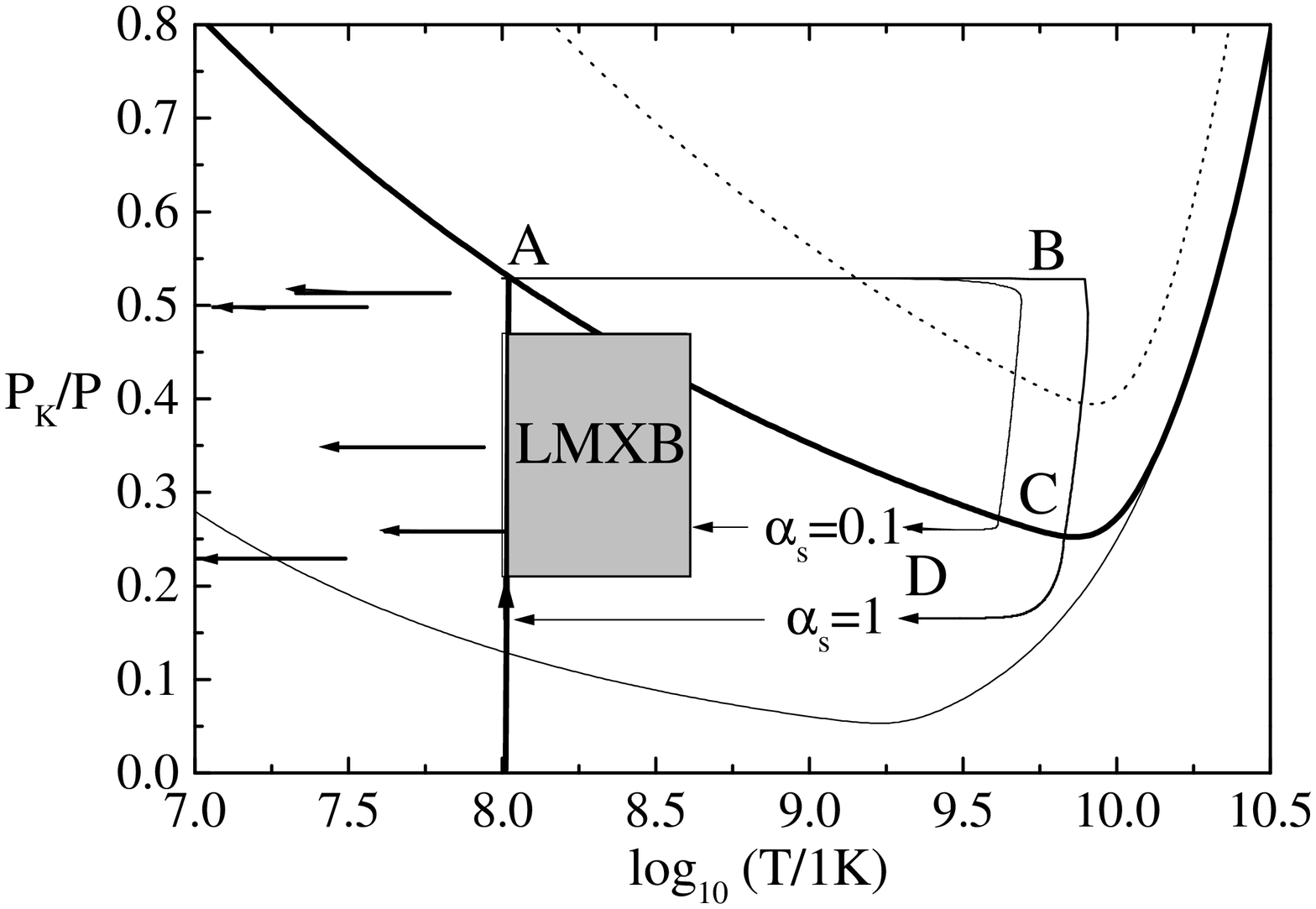}}
%\vskip -0.15in
\figcaption{\label{fig1}
The r-mode instability window relevant for
  old neutron stars. We show results for the simplest (crust-free)
  model, where gravitational radiation reaction is balanced by
  standard shear viscosity at low temperatures and bulk viscosity at
  high temperatures (thin solid line).  Also shown are Bildsten and
  Ushomirsky's estimate for a star with a crust (dashed line) and our
  improved estimate of this situation (thick solid line).  The r-mode
  is potentially unstable in the region above each curve. The
  illustrated results correspond to a canonical neutron star for which
  mass shedding at the equator sets in at the Kepler period
  $P_K\approx 0.8$~ms.  We illustrate two typical r-mode cycles (for
  mode saturation amplitudes $\alpha_s=0.1$ and 1), resulting from
  thermo-gravitational runaway after the onset of instability.  Once
  accretion has spun the star up to the critical period (along the
  indicated spin-up line (thick vertical line)) the r-mode becomes
  unstable and the star evolves along the path A-B-C-D. After a month
  or so, the mode is stable and the star will cool down until it again
  reaches the spin-up line.  For comparison with observational data,
  we indicate the possible range of spin-periods inferred from current
  LMXB data (shaded box) as well as the observed periods and estimated
  upper limits of the temperature (cf. \citet{akst99}) of some of the
  most rapidly spinning MSP (short arrows).}

\end{minipage}
\hfill
\begin{minipage}[t]{3.5in}
\epsfysize=6cm
\centerline{\epsfbox{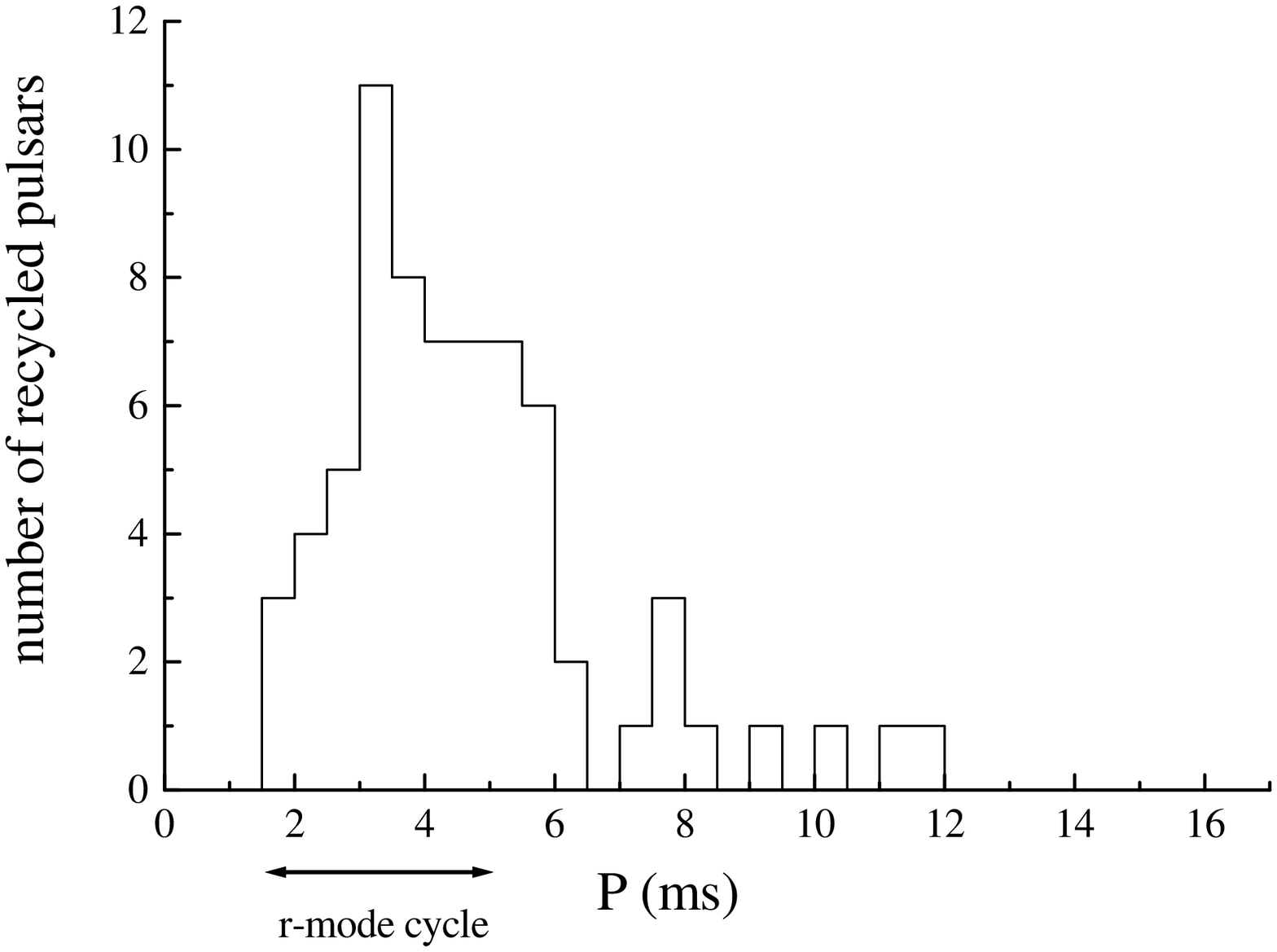}}
\vskip -0.15in
\figcaption{\label{fig2} We compare the ``r-mode cycle'' predicted
  for accreting neutron stars to the observed rapidly spinning
  pulsars.  The observed data is taken from the Princeton pulsar
  database (http://pulsar.princeton.edu/pulsar/catalog.shtml) as well
  as a sample of MSP recently discovered in 47Tuc
  \citep{Camilo}.}

\end{minipage}\end{figure*}

The fact that our revised instability curve for r-modes damped by
dissipation in a viscous boundary layer agrees well with the fastest
observed neutron star spin frequencies, cf. Figure~\ref{fig1},
motivates us to speculate further on the relevance of the instability.
We want to model how the potential presence of an unstable r-mode
affects the spin-evolution of rapidly spinning, accreting neutron
stars. To do this we use the phenomenological two-parameter model
devised by \citet{owen98}, which is centered on evolution equations
for the rotation frequency $\Omega$ and the (dimensionless) r-mode
amplitude $\alpha$.  Complete details of our particular version of
this model will be given elsewhere.

At the qualitative level, our results are not surprising.  Accreting
stars in the LMXB are expected to have core temperatures in the range
$1-4\times10^8$~K \citep{brown}. For such temperatures the dissipation
due to the viscous boundary layer gets weaker as the temperature
increases. Consequently, the situation here is essentially identical
to that considered by \citet{levin99} (see also \citet{spruit} and
\citet{bildsten99}).  After accreting and spinning up for something
like $10^7$ years, the star reaches the period at which the r-mode
instability sets in. For our particular estimates this corresponds to
a period of 1.5~ms (at a core temperature of $10^8$~K). It is notable
that this value is close to the 1.56~ms period of PSR1937+21. Once the
r-mode becomes unstable (point A in Figure~\ref{fig1}), viscous
heating (now mainly due to the energy released in the viscous boundary
layer) rapidly heats the star up to a few times $10^9$~K. The r-mode
amplitude increases until it reaches a prescribed saturation level
(amplitude $\alpha_s$) at which unspecified nonlinear effects halt
further growth (point B in Figure~\ref{fig1}).  Once the mode has
saturated, the neutron star rapidly spins down as excess angular
momentum is radiated as gravitational waves.  When the star has spun
down to the point where the mode again becomes stable (point C in
Figure~\ref{fig1}), the amplitude starts to decay and the mode plays
no further role in the spin evolution of the star (point D in
Figure~\ref{fig1}) unless the star is again spun up to the instability
limit.  Two examples of such r-mode cycles (corresponding to
$\alpha_s=0.1$ and 1, respectively) are shown in Figure~\ref{fig1}.

The real surprise here concerns the quantitative predictions of our
model.  As already mentioned, the model suggests that an accreting
star will not spin up beyond 1.5~ms. This value obviously depends on
the chosen stellar model, but it is independent of the r-mode
saturation amplitude and only weakly dependent on the accretion rate
(through a slight change in core temperature).  In fact, the accretion
rate only affects the time it takes the star to complete one full
cycle. As soon as the mode becomes unstable the spin-evolution is
dominated by gravitational radiation and viscous heating. Once the
star has gone through the brief phase when the r-mode is active it has
spun down to a period in the range 2.8-4.8~ms (corresponding to
$0.01\leq \alpha_s \leq 1$).  Based on these results we propose the
following spin-evolution scenario: An accreting neutron star will
never spin up beyond (say) 1.5~ms. Once it has reached this level the
r-mode instability sets in and spins the star down to a several ms
period. At this point the mode is again stable and continued accretion
may resume to spin the star up.  Since the star must accrete roughly
$0.1M_\odot$ to reach the instability point, and the LMXB companions
have masses in the range $0.1-0.4M_\odot$, it can pass through several
``r-mode cycles'' during its lifetime.

Let us confront this simple model with current observations. To do
this we note that our model leads to one main prediction: Once an
accreting neutron star has been spun up beyond (say) 5~ms it must
remain in the rather narrow range of periods $1.5-5$~ms until it has
stopped accreting and magnetic dipole braking eventually slows it
down.  Since a given star can go through several r-mode cycles before
accretion is halted one would expect most neutron stars in LMXB and
the MSP to be found in the predicted range of rotation rates.  As is
clear from Figure~\ref{fig1}, this prediction agrees well with the
range of rotation periods inferred from observed kHz quasiperiodic
oscillations in LMXB.  The observed range shown in Figure~\ref{fig1}
corresponds to rotation frequencies in the range 260-590~Hz (cf.
\citet{vanderklis00}). Our model also agrees with the observed data
for MSP, which are mainly found in the range $1.56-6$~ms, see
Figure~\ref{fig2}.  In other words, our proposed model is in agreement
with current observed data for rapidly rotating neutron stars.
 
Finally, it is worthwhile discussing briefly the detectability of the
gravitational waves that are radiated during the relatively short time
when the r-mode is saturated and the star spins down. As was argued by
\citet{levin99}, the fact that the r-mode is active only for a small
fraction of the lifetime of the system (something like 1 month out of
the $10^7$~years it takes to complete one full cycle) means that even
though these sources would be supremely detectable from within our
galaxy the event rate is far too low to make them relevant. However,
it is interesting to note that the spin-evolution is rather similar to
that of a hot young neutron star once the r-mode has reached its
saturation amplitude.  This means that we can analyze the
detectability of the emerging gravitational waves using the framework
of \citet{owen98}.  We then find that these events can be observed
from rather distant galaxies. For a source in the Virgo cluster
(assumed to be at a distance of 15~Mpc) these gravitational waves
could be detected with a signal to noise ratio of a few using LIGO~II.
However, even at the distance of the Virgo cluster these events would
be quite rare.  By combining a birth rate for LMXB of $7\times
10^{-6}$ per year per galaxy with the fact that the the volume of
space out to the Virgo cluster contains $\sim 10^3$ galaxies, and the
possibility that each LMXB passes through (say) four r-mode cycles
during its lifetime we deduce that one can only hope to see a few
events per century in Virgo. In order to see several events per year
the detector must be sensitive enough to detect these gravitational
waves from (say) 150~Mpc. This would require a more advanced detector
configuration such as a narrow-banded LIGO~II.  We will discuss this
issue in more detail elsewhere.

\section{Additional remarks}

Before concluding our discussion, we recall that the initial
excitement over the r-mode instability was related to the fact that it
provided an explanation for the relatively slow inferred spin rates
for young pulsars. In view of this, it is natural to digress somewhat
and discuss how the picture of the r-mode instability in hot, newly
born neutron stars is affected by the possible formation of a solid
crust.  Hence, we consider the evolution of a neutron star just after
its birth in a supernova explosion.  At a first glance, we might
expect to model its r-mode amplitude in the standard way, cf.
\citet{owen98}, using the normal (crust-free) fluid viscous damping
times for stellar temperatures above the melting temperature of the
crust ($T_m$), and the viscous boundary layer damping time for
temperatures below $T_m$.  However, the situation is a little more
complicated than this.  Recall that the latent heat (i.e. the Coulomb
binding energy) of a typical crust is $E_{lat} \sim 10^{48}$~ergs,
while the r-mode energy is $E_m\approx 2 \alpha^2 (\mbox{1
ms}/P)^2\times 10^{51}$~ergs.  Provided that the time taken for the
star to cool to $T_m$ is sufficiently long, the energy in the r-mode
(which grows exponentially on a timescale $t_{gw} \sim 20$~s) will
exceed $E_{lat}$, preventing the formation of the crust, even when $T
< T_m$.  Then the star will spin down in the manner described by, eg.
\citet{owen98}.  This phase will end either because the star leaves
the instability region of the $\Omega-T$ plot (see, for example,
Fig.~1 of Owen et al.), or because the mode energy in the outer layers
of the star (where the crust is going to form) has fallen below the
crustal binding energy. We can estimate that this would happen at a
frequency $\approx 70\mbox{ Hz}/\alpha_s$, by equating $E_{lat}$ to
roughly 10 \% of $E_m$). A more accurate treatment would take into
account the \emph{local} kinetic energy of fluid elements and since
the latter is smaller near the poles than near the equator, the crust
might form earlier at the poles. Clearly the problem of crust
formation in an oscillating star requires further investigation.  The
final spin period will be around 15~ms, if the r-mode grows to an
amplitude of $\alpha_s \sim 1$, consistent with the extrapolated
initial spin rates of many young pulsars. On the other hand, if the
mode is not given time to grow very large it will not prevent crust
formation at $T_m$.  Such a scenario was described by
\citet{bildsten99}, who noted that the r-mode instability would then
not spin the star down beyond a much higher frequency. Using our
estimated timescales the resultant spin period would be $3-5$~ms in
this scenario.  Which scenario applies depends sensitively on the
early cooling of the star, the crustal formation temperature and
perhaps most importantly the initial amplitude of the r-mode following
the collapse. It is, in fact, possible that both routes are viable and
that a bimodal distribution of initial spin periods results. A likely
key parameter is whether the supernova collapse leads to a large
initial r-mode amplitude $\sim \alpha_s$ or not.  An initial period of
$\sim 15$~ms would fit the long established data for the Crab, while
the recently discovered 16~ms pulsar PRS~J0537-6910 \citep{marshall}
requires a considerably shorter initial period of a few ms.

In conclusion, we have reexamined the effect that the dissipation due
to a possible viscous boundary layer in a neutron star with a solid
crust has on the stability of the r-modes.  By combining our new
estimates with the thermal runaway introduced by \citet{levin99} and
\citet{spruit}, we arrive at a spin-evolution model that agrees with
present observations for rapidly spinning neutron stars.  In
particular, our predictions agree well with observations of both LMXB
and MSP. Furthermore, the model can potentially explain the
extrapolated spin periods of the young pulsars. Since it brings out
this unified picture, our simple model has many attractive features,
and we are currently investigating it in more detail.

\acknowledgements We thank L. Bildsten, W. Kluzniak, B. Sathyaprakash,
H. Spruit and G. Ushomirsky for comments on a draft version of this
paper. This work was supported by PPARC grant PPA/G/1998/00606 to NA.

\end{document}